\documentstyle[12pt,psfig]{article}
\newcommand{\s}{\rm}

\newcommand{\be}{\begin{equation}}
\newcommand{\ee}{\end{equation}}

\newcommand{\bea}{\begin{eqnarray}}
\newcommand{\eea}{\end{eqnarray}}
\newcommand{\bef}{\begin{figure}}
\newcommand{\eef}{\end{figure}}

\begin{document}
\begin{flushright}
{\normalsize
January 2002}
\end{flushright}
\vskip 0.2in
\begin{center}

{\Large{\bf Space-time evolution of
ultra-relativistic heavy ion collisions and hadronic spectra}}
\end{center}
\vskip 0.1in
\begin{center}
B. K. Patra$^a$\footnote{Present Address: Saha Institute of
Nuclear Physics, 1/AF Bidhan Nagar, Kolkata 700 064}, 
Jan-e Alam$^a$, Pradip Roy$^b$,
 Sourav Sarkar$^a$
 and Bikash Sinha$^{a,b}$
\end{center}
\noindent
{\small {\it a) Variable Energy Cyclotron Centre,
     1/AF Bidhan Nagar, Kolkata 700 064, India}}
\vskip 0.1in
\noindent
{\small {\it b) Saha Institute of Nuclear Physics,
           1/AF Bidhan Nagar, Kolkata 700 064,
           India}}

\addtolength{\baselineskip}{0.4\baselineskip} %wide line spacing

%\date{today}
\parindent=20pt
\vskip 0.1in
\begin{abstract}
The space-time evolution of the hot and dense matter formed after 
the collisions of heavy nuclei at ultra-relativistic energies
is investigated using (3+1) dimensional hydrodynamical models.
The effects of the spectral shift of the hadronic properties
are incorporated in the equation of state (EOS) of the evolving matter. 
In-medium shift of hadronic properties are considered for
Quantum Hadrodynamics (QHD) and universal scaling scenarios.
It is found that the EOS for the hadronic matter
for universal scaling of hadronic masses (except
pseudoscalar) is similar to the recent lattice results.
We observe that the space-time volume of the hadronic
matter at the freeze-out is considerably different from the one
when medium effects on the hadrons are ignored. 
The sensitivity of the results on the initial radial
velocity profile is investigated.
The transverse mass spectra of pions and protons
of NA49 collaboration are analyzed. 

\end{abstract}

\section{Introduction}
One of the main motivations to study the nucleus-nucleus
collisions at ultra-relativistic energies is to create a
very hot and dense system of strongly interacting matter,
a situation conducive for the formation of 
Quark Gluon Plasma (QGP)~\cite{qm01}. 
It is not easy, however, to obtain information
about the early stage of this matter 
because of the very short lifetime of the
dense system. Electromagnetically interacting
particles (photons and lepton pairs)
are considered to be the ideal probes for the
early state of the matter.
On the other hand, observations through the transverse 
mass ($m_T$) spectrum of the  hadrons
provide valuable informations about the situation when the
thermal system freezes out~\cite{hadpt1,hadpt2,hadpt3,hadpt4,hadpt5}, 
{\it i.e.} the stage
when the system disassembles to individual hadrons.
Thus the study of the hadronic spectra gives
snapshots of the later stages of the nuclear collision
dynamics.  

It has been emphasized recently
that the  properties of the hadrons will be modified
due to its interactions with the particles in the
thermal bath. It is also worth emphasizing here
that as yet it has not been possible to explain
the enhanced dilepton production measured by the 
CERES/NA45 collaborations ~\cite{ceres} in the low 
invariant mass region (below the $\rho$-peak)
without the in-medium modifications of
the vector mesons~\cite{RW}. We have recently analyzed ~\cite{jarap}
the WA98 photon data~\cite{wa98} by incorporating the 
in-medium modifications of hadrons. In the light of
these observations we study the hadronic spectra
in a scenario where the hydrodynamic evolution  
contains the in-medium modifications of hadrons
through the parametrization of the equation of state (EOS).
In particular we will show here that the NA49~\cite{na49}
 hadronic  spectra can 
be explained by the same initial and freeze-out 
conditions with the EOS which explain the photon and dilepton
data at CERN SPS energies.
We consider two possible scenarios: (i) nucleus + nucleus
$\longrightarrow$ QGP $\longrightarrow$ hadrons  
and (ii) nucleus + nucleus 
$\longrightarrow$ excited hadronic matter $\longrightarrow$ hadrons
(but the properties of hadrons
in the thermal bath are different from the vacuum
due to its interaction with the particles in the thermal bath),
to investigate the evolution of the matter formed after
the collisions.

In the next section we review the hydrodynamical model
and the EOS used in the prersent work. In section 3
we present the results of our calculations and 
section 4 is devoted to summary and discussions.

\section{The Single-particle Spectra in the Hydrodynamical Model}

We will assume that the produced matter reaches a state
of local thermodynamic equilibrium after a proper time,
$\tau_i\,\sim 1$ fm/c~\cite{jdb}. 
The system continues to expand in space and time till the freeze-out
undergoing a phase transition to the hadronic matter in 
the process, if QGP is formed initially. 
The conversion of thermal energy of the system
into the collective flow continues as long as 
collisions are sufficiently frequent. 
Eventually the density of the particles decreases and correspondingly their
mean free path increases. 
When the mean free path becomes of the order
of the size of the system, hadrons act as non-interacting free particles.
The detailed description of this decoupling
transition, or freeze-out, is complicated. A simple algorithm is obtained if
one assumes that the break-up occurs when the temperature drops
down to a given decoupling temperature $T_F$ ; this temperature, which
is typically of the order of the pion mass, defines a space-time
surface $T(r,t) = T_F$, usually referred to as the freeze-out surface (FS),
on which the system ceases to behave collectively.
A fluid element which crosses this
surface liberates particles which, in the rest frame of the fluid
element, have a thermal distribution at a temperature $T_F$. Knowing
the velocity of the fluid element on the decoupling surface, one can then
determine the momentum distribution of the final particles.
The observed
momentum distribution will thus be characterized by the temperature
and collective velocity of the system at the FS.
For given initial conditions and EOS we solve the hydrodynamical equations 
to obtain the FS, determined by the condition,
$T(\tau,r)=T_F$. 

The expansion of the system is described by the energy-momentum
conservation of an ideal fluid :
\be
\partial_\mu\,T^{\mu \nu} =0;\,\,\,\,\, 
\quad T^{\mu\nu}=(\epsilon+P)u^\mu\,u^\nu+g^{\mu\nu}P 
\label{hydro}
\ee
where $\epsilon$ is the energy density, $P$ is the pressure measured in the
frame co-moving with the fluid, and $u^\mu$ is the fluid four-velocity.

With the assumption that
the system undergoes a boost-invariant longitudinal expansion along the $z$
axis~\cite{jdb} and a cylindrically symmetric transverse expansion 
the hydrodynamic Eqs.~\ref{hydro} reduce to
\be
 \partial_{\tau}T^{00}+\frac{1}{r}\partial_r(rT^{01})+
\frac{1}{\tau}(T^{00}+P)=0
\label{hydro1}
\ee
 and
\be
\partial_{\tau}T^{01}+
\frac{1}{r}\partial_r\left[r(T^{00}+P)v_r^2\right]+
\frac{1}{\tau}T^{01}+\partial_rP=0
\label{hydro2}
\ee
where $u^{\mu}=\gamma(t/\tau, \vec{v_r},z/\tau)$

Eqs.~(\ref{hydro1}) and (\ref{hydro2}) are solved by the 
relativistic version of 
the flux corrected transport algorithm~\cite{hvg}
with the following initial energy density profile,
\be
\epsilon(\tau_i,r)=\frac{\epsilon_0}
{1+\exp\left[(r-R_A)/\delta\right]}.
\label{e0}
\ee
where  $R_ A$ is the nuclear radius and $\delta$ is the surface thickness 
parameter. We have taken $\delta =1/2$ in our calculations.
We will study the sensitivity of the flow with
the following two kinds of initial velocity profile~\cite{heinz,pbm},
\be
 v_r(\tau_i,r)=v_0\frac{r}{R_A}
\label{v01}
\ee
and ~\cite{hvg}
\be
v_r(\tau_i,r)=v_0\,\left[1-\frac{1}
{1+\exp\left[(r-R_A)/\delta\right]}
\right]
\label{v02}
\ee
The FS, which is required
as an input to evaluate the single particle spectra
is obtained by solving the Eqs.~(\ref{hydro1}) and (\ref{hydro2})
with the initial conditions~(\ref{e0}) and ~(\ref{v01}) (or eq.\ref{v02}) 
for a given EOS. 

The single particle transverse momentum distribution 
in the hydrodynamical model is  given by the well-known
Cooper-Frye formula~\cite{cooper},
\be
E{dN\over  d^3p}  =  {g\over  (2\pi)^3}\int_{\sigma}  f(r,p)p^\mu
d\sigma_\mu,
\ee
where $f(r,p)$ is  the Boltzmann distribution (quantum statistical
effects are neglected here)
\be
f(r,p) = \exp [-(p.u-\mu)/T].
\ee
$T$, $\mu$ and $u^\mu$ are the temperature,
chemical potential  and
four-velocity of the fire-cylinder respectively. 
The three dimensional space-time surface $\sigma$,
whose surface elements are specified in terms of the
four vector $d\sigma^\mu=(d^3r,\,dtdS)$, is determined
by the freeze-out condition, $T(\tau,r)=T_F$ as mentioned
before.  For a system with cylindrical symmetry and boost invariance
along the longitudinal direction the single particle
momentum ($p_T$) distribution is given by~\cite{heinz,ruuskanen,blaizot},
\begin{eqnarray}
\left( {dN\over dy p_Tdp_T}\right)_{y=0}
&=& {g\over \pi} \int_\sigma r~dr~\tau_F(r)    \nonumber\\
& & \left\{ m_tI_0 \left( {p_T\sinh y_t\over T} \right)\right.
          K_1 \left( {m_t\cosh y_t\over T} \right) \nonumber \\
& & -\left( {\partial\tau_F\over\partial r} \right) p_T
          I_1 \left( {p_T\sinh y_t\over T} \right)
    \left. K_0 \left( {m_t\cosh y_t\over T} \right) \right\}
\end{eqnarray}
where $\tau_F(r)$ refers to the freeze-out time which, in general,
 depends on $r$ and $y_t$ is the transverse rapidity. 

\subsection{Equation of state}

The set of hydrodynamic eqs.~(\ref{hydro1}) and (\ref{hydro2}) are not 
closed by itself;
the number of unknown variables exceeds the number of equations by one.
Thus a functional relation between any two  
variables is required so that the system becomes deterministic. 
The most natural course is to look for
such a relation between the pressure $P$ and the energy density $\epsilon$. 
Under the assumption of local thermal equilibrium, this functional relation
between $P$ and $\epsilon$ is the EOS.
Obviously, different EOS's will govern the hydrodynamic
flow quite differently and as far as the search 
for QGP is concerned, the goal is to look for
distinctions in the observables due to the different EOS's (corresponding
to the novel state of QGP vis-a-vis that for the hadronic matter). 
It is thus imperative to understand in what respects the two EOS's differ 
and how they affect the evolution in space and time. 

A physically intuitive way of understanding the role of the EOS in governing
the hydrodynamic flow lies in the fact that the velocity of sound $c_{s}^{2}
=(\partial P/\partial\epsilon)_s$ sets an intrinsic scale in the hydrodynamic
evolution. One can thus write a simple parametric form for the EOS:
$P=c_{s}^{2}(T)\epsilon$. 
Clearly, inclusions of interactions may drastically alter the value of 
$c_{s}^{2}$.
In the present work MIT bag model equation of state
is assumed for the QGP where the energy density and 
pressure are given by
\be
\epsilon_Q=g_Q{\pi^2\over 30}T^4+B,
\ee
and
\be
P_Q=g_Q\frac{\pi^2}{90}T^4-B.
\ee
The effective degrees of freedom in QGP, $g_Q=37$ for two flavours.
The entropy density $s_Q$ is given by $s_Q=2g_Q(\pi^2/45)T^3$. 

As  the  hydrodynamic expansion starts, the QGP begins to cool
until  the  temperature drops down to the critical  
temperature  $T_c$.
At  this
instant, the phase transition  to  the  hadronic  matter  starts.
Assuming  that  the  phase transition is a first-order one, the 
released latent heat maintains the temperature  of
the  system  at  the  critical temperature $T_c$, even though the
system continues to expand. The cooling due to expansion is compensated
by the latent heat liberated during the process.
We neglect  the  scenarios  of  supercooling or
superheating and any other plausible explosive events. 
This process continues until all the matter has
converted to the hadronic phase with the
temperature remaining constant at $T=T_c$. From then on,
the system continues to expand, governed by  the  EOS  of  the  hot
hadronic  matter  till the freeze-out temperature $T_F$
at the proper time $\tau_F$. Thus the
appearance of the so called mixed phase at $T=T_c$, when QGP  and
hadronic  matter  co-exist,  is a direct consequence of the first
order phase transition. Apart from the role in  QGP  diagnostics,
the  possibility of the mixed phase affects the bulk
features of the evolution process also. 

In the hadronic phase 
one must consider the effect of the presence
of heavier particles and the change in their masses due to finite temperature
effects.
The ideal limit of treating the hot hadronic matter as a gas of pions 
originated from the expectation that in the framework of local 
thermalization the system would be dominated by the lowest mass hadrons 
while the higher mass resonances would be
Boltzmann suppressed. 
Indirect justification of this assumption comes from the experimental
observation in high energy collisions that most of the secondaries are pions.
Nevertheless, the temperature of the system is 
higher than
$m_{\pi}$ during a major part of the evolution and at these temperatures the
suppression of the higher mass resonances may not be complete. It may 
therefore be more realistic to include higher mass resonances in the 
hadronic sector, their relative
abundances being governed by the condition of (assumed) thermodynamic 
equilibrium.
We assume that 
the hadronic phase consists of $\pi$, $\rho$, $\omega$, $\eta$, $a_1$  
mesons and nucleons. The nucleons and heavier mesons are expected 
to play an important role in the EOS particularly,
in a scenario where mass of the 
hadrons decreases with temperature. 

The energy density and pressure
for such a system of hadrons are given by
\be
\epsilon_H=\sum_{h={\s mesons}} \frac{g_h}{(2\pi)^3} 
\int d^3p\,E_h\,f_{BE}(E_h,T)
+\frac{g_N}{(2\pi)^3} 
\int d^3p\,E_N\,f_{FD}(E_N,T)
\ee
and
\be
P_H=\sum_{h={\s mesons}} \frac{g_h}{(2\pi)^3} 
\int d^3p\frac{p^2}{3\,E_h}f_{BE}(E_h,T)
+\frac{g_N}{(2\pi)^3} 
\int d^3p\frac{p^2}{3\,E_N}f_{FD}(E_N,T)
\ee
where the sum is over all the mesons under consideration and $N$ stands
for nucleons. The in-medium mass of the hadrons ($H$)
enters through the relation, $E_H=\sqrt{p^2 + m_H^{\ast 2}}$
~~(the asterisk denotes the effective mass in the medium).          
The entropy density is then given by
\be
s_H=\frac{\epsilon_H+P_H}{T}\,\equiv\,4a_{\s{eff}}(T)\,T^3
= 4\frac{\pi^2}{90} g_{\s{eff}}(m^\ast(T),T)T^3
\label{entro}
\ee
where  $g_{\s eff}$ is the effective statistical degeneracy.
Thus, we can visualize the finite mass of the hadrons
having an effective degeneracy $g_{\s{eff}}(m^\ast(T),T)$. 
We consider the effects of in-medium  mass 
on the EOS both for the universal scaling
scenario~\cite{brpr} and 
the Quantum Hadrodynamic (QHD) scenario. 
According to the universal scaling scenario
the hadronic masses (except
pseudoscalars) approaching zero 
near the critical temperature as follows~\cite{brpr},
\be
m_H^\ast/m_H=(1-T^2/T_c^2)^{\lambda}
\label{emass}
\ee
Results for various values of the exponent $\lambda$ 
will be discussed below.
For a detailed discussions on the in-medium masses of hadrons in QHD
interaction we refer to ~\cite{hsk,npa12,sxk}. 
The temperature dependence of the mass of nucleon ($m_N^\ast$),
$\rho$ ($m_\rho^\ast$) and $\omega$ ($m_\omega^\ast$) in QHD
are parametrized as follows:
\be
m_N^\ast=m_N\left[1-0.0264\left(\frac{T ({\s GeV})}{0.16}\right)^{8.94}\right].
\ee
\be
m_\rho^\ast=m_\rho\left[1-
0.127\left(\frac{T({\s GeV})}{0.16}\right)^{5.24}\right]
\ee
\be
m_\omega^\ast=m_\omega\left[1-
0.0438\left(\frac{T({\s GeV})}{0.16}\right)^{7.09}\right].
\ee
The effective mass of the $a_1$ meson, 
$m_{a_1}^\ast$ has been estimated from $m_{\rho}^\ast$ by using Weinberg's
sum rule~\cite{weinberg}.

\section{Results}

In Fig.~\ref{fig1} the variation of the effective
degeneracy (the co-efficient of $\epsilon/T^4$ differs
from the effective degeneracy by a multiplicative factor 
of $\pi^2/30$) is plotted
as a function of temperature. The rapid increase of the
effective degeneracy near the critical
temperature as per lattice QCD calculations~\cite{lattice}
is reasonably well described 
when the effective masses vary according to the eq.(\ref{emass})
(see also~\cite{hb}).

%%%%%%%%%%%%%% Fig. 1 %%%%%%%%%%%%%%%%%%%%%%%%%%%%%
\bef
\centerline{\psfig{figure=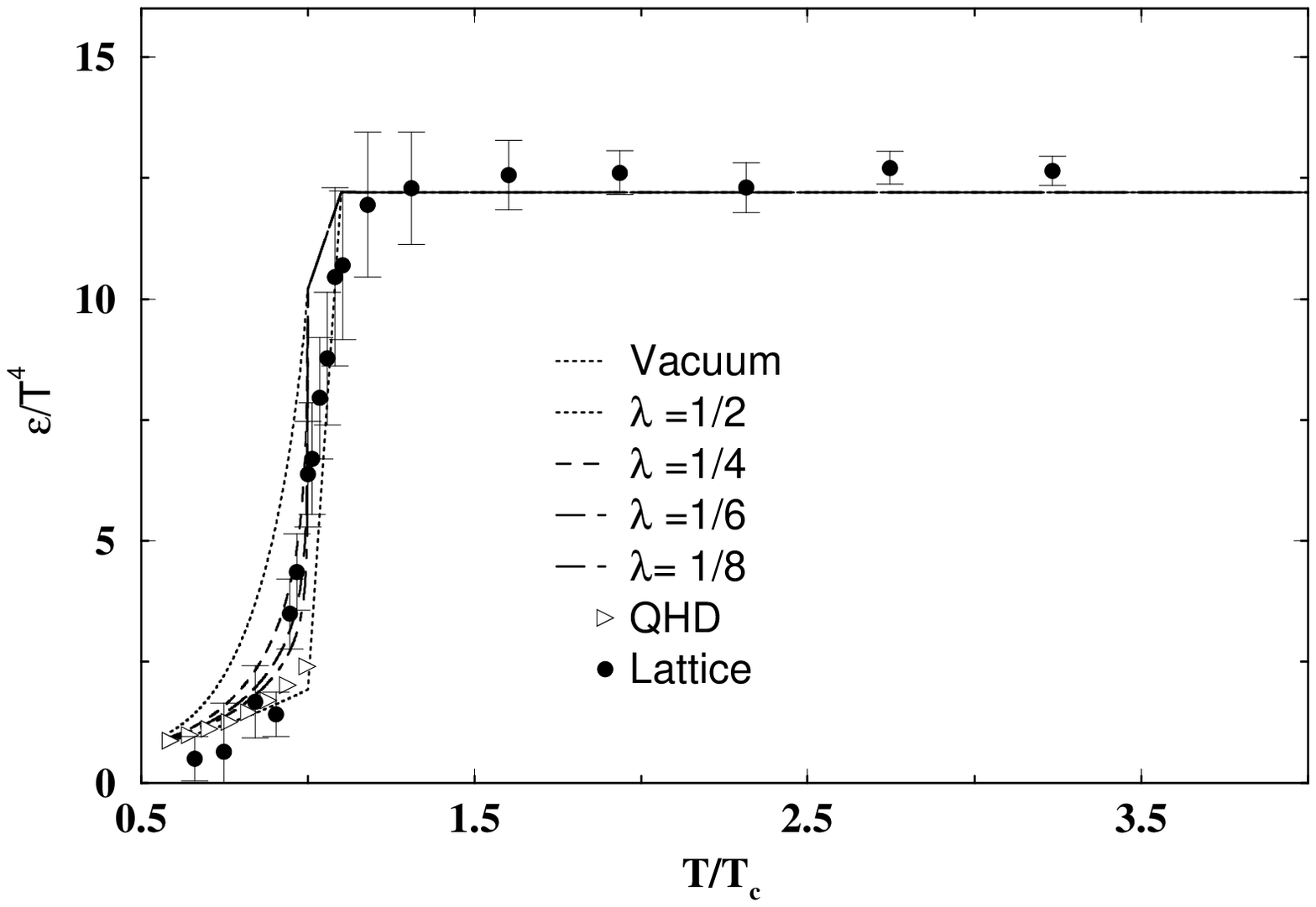,height=5.5cm,width=8cm}}
\caption{The energy density $\epsilon$ in
unit of $T^4$ for various equations of state used
in the present work are plotted
as a function of temperature in unit 
of the critical temperature, $T_c$. The 
filled circles denote the lattice results~\protect{\cite{lattice}}.
The dashed line indicates results
with vacuum masses.
}
\label{fig1}
\eef
%%%%%%%%%% End of Fig. 1 %%%%%%%%%%%%%%%%%%%%%%%%%%%%%%%%

We solve the following equation to estimate the initial 
temperature,
\be
\frac{dN_\pi}{dy}=
\frac{45\zeta(3)}{2\pi^4}\pi\,R_A^2 4a_{\s{eff}}(T_i)T_i^3\tau_i
\label{dndy}
\ee
where
$a_{\s{eff}}(T_i)=({\pi^2}/{90})\,g_{\s{eff}}(m^\ast(T_i),T_i)$.
The change in the expansion dynamics
as well as the value of the initial temperature due
to medium effects enters the calculation 
through the effective statistical degeneracy.

%%%%%%%%%%%%%%%%%%%%% Table 1 %%%%%%%%%%%%%%%% 
\renewcommand{\arraystretch}{1.5}
\vskip 0.2in
\begin{center}
\begin{tabular}{|c|c|c|c|c|c|}
\hline
& QGP & vacuum & QHD &  Scaling \\
&     & mass   &     & $T_c$=200 MeV \\
\hline
$dN/dy$ & $T_i$ (MeV) & $T_i$ (MeV) & $T_i$ (MeV) &  $T_i$ (MeV) \\
\hline
700 & 196 & 245 & 220 & 205 \\
\hline
\end{tabular}
\end{center}
%%%%%%%%%%%%%%%%%%%%% end Table 1 %%%%%%%%%%%%%%%% 

\noindent{ Table I: Initial temperatures for QGP and
hadronic initial states.}

We consider Pb + Pb collisions at CERN SPS energies. 
Taking $dN_{\pi}/dy=700$
for Pb + Pb collisions and 
$\tau_i=1$ fm/c,
Eqs.~(\ref{entro}) and ~(\ref{dndy}) are solved self consistently
to get the initial temperatures for different 
mass variation scenarios shown in Table I.
For a given $dN/dy$ and $\tau_i$ the value of $T_i$
is higher in case of vacuum masses (smaller $a_{eff}$)
compare to the universal scaling scenario (higher $a_{eff}$).

\begin{figure}
\begin{tabular}{cc}         
   \psfig{file=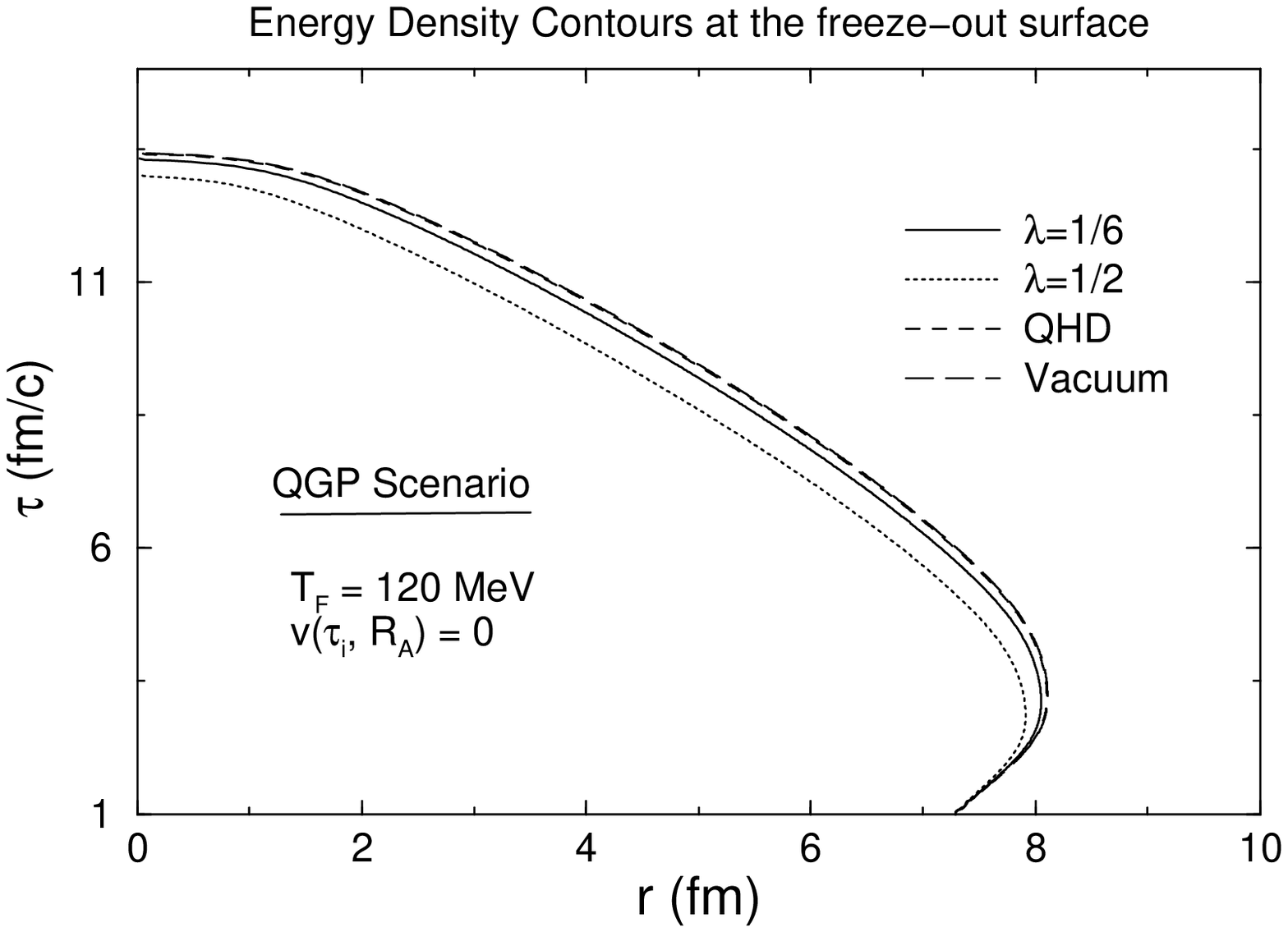,width=7.5cm}  &
   \psfig{file=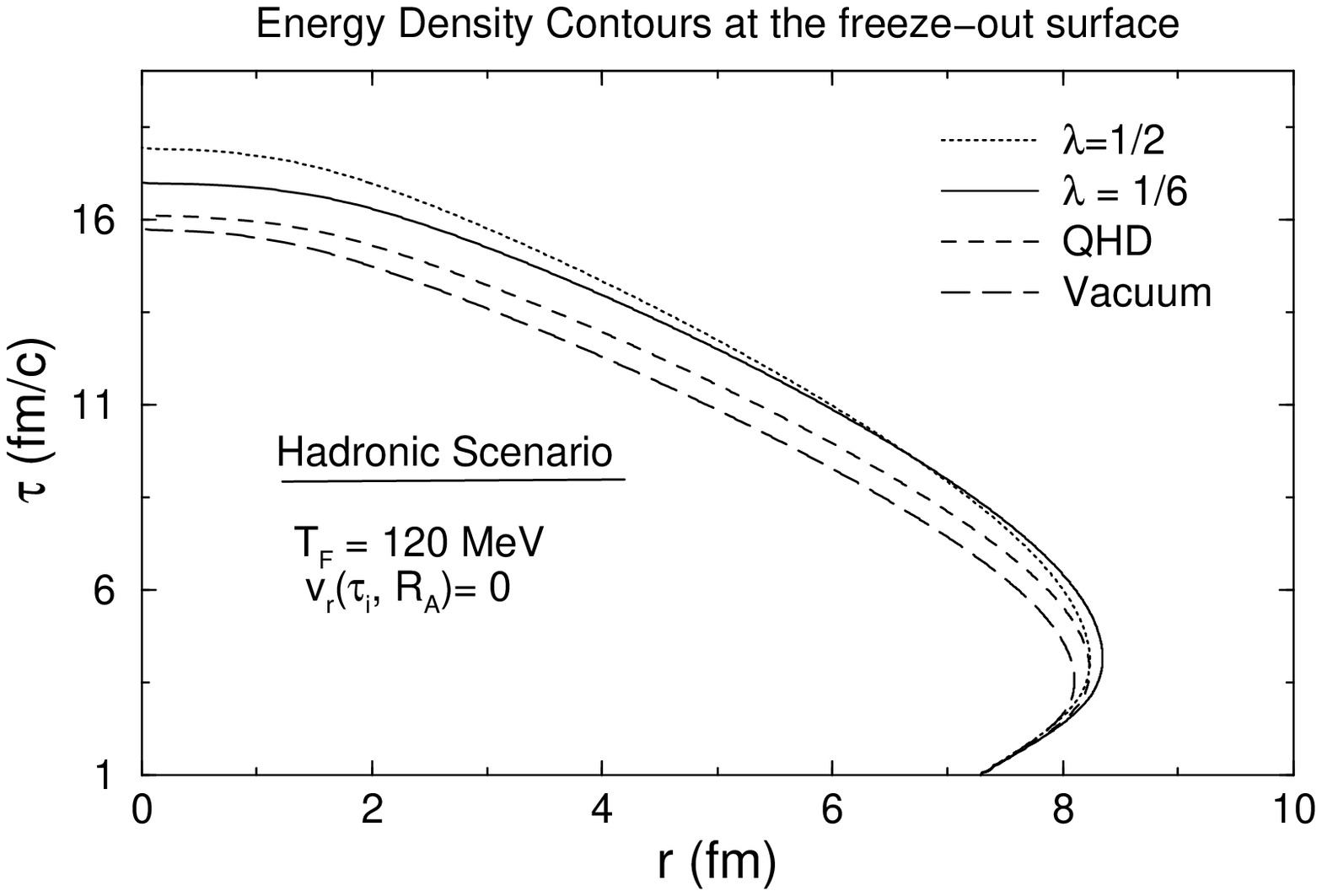,width=7.5cm}  \\
\end{tabular}
\caption{Constant energy density contours at the  FS
(in $r-\tau$ plane) for
initial velocity profile of Eq.~\protect{\ref{v01}} with $v_0=0$.
Left (Right) : QGP (hadronic) initial state.}
\label{fig1ab}
\end{figure}

Next we consider the constant energy density contours
at the FS.  
The value of $T_F$ is taken as $120$ MeV in the present calculations.
Eqs.~(\ref{e0}) and (\ref{v01}) are used  for the initial 
energy density and velocity profiles respectively.
In case of QGP formation the initial energy density is 2.6 
GeV/fm$^3$ corresponding to an initial temperature of 196 MeV.
In this case the in-medium effects 
on the hadronic phase formed after the phase transition 
is taken into account.
The energy density at the freeze-out point, $\epsilon_F$
$(\propto a_{eff}(T_F)T_F^4)$ is different for different 
hadronic model. $\epsilon_F$ is larger for the universal
scaling scenario with $\lambda=1/2$  compared to the 
case of the hadronic gas with vacuum masses, because for a given $T_F$,
$a_{eff}$ is larger in the former case. As a consequence of this
the FS is smaller in universal scaling scenario
compared to the vacuum scenario. This is demonstrated 
in the left panel of fig.~\ref{fig1ab}. The FS corresponding to
QHD and the vacuum are indistinguishable because of
the small medium effects in QHD especially at the late stage 
of the evolution, where the temperature is lower. 
In the right panel of fig.~\ref{fig1ab} 
we display the constant $\epsilon_F$
surfaces for hadronic initial
state.  The velocity of sound in case of universal scaling is
smaller than the scenario when medium effects are ignored. 
As a result the FS for the universal scaling scenario
is larger (for a given $T_F$) as compared to the vacuum
mass scenario, because of the slower cooling of the system
in the former case.  

\begin{figure}
\begin{tabular}{cc}         
   \psfig{file=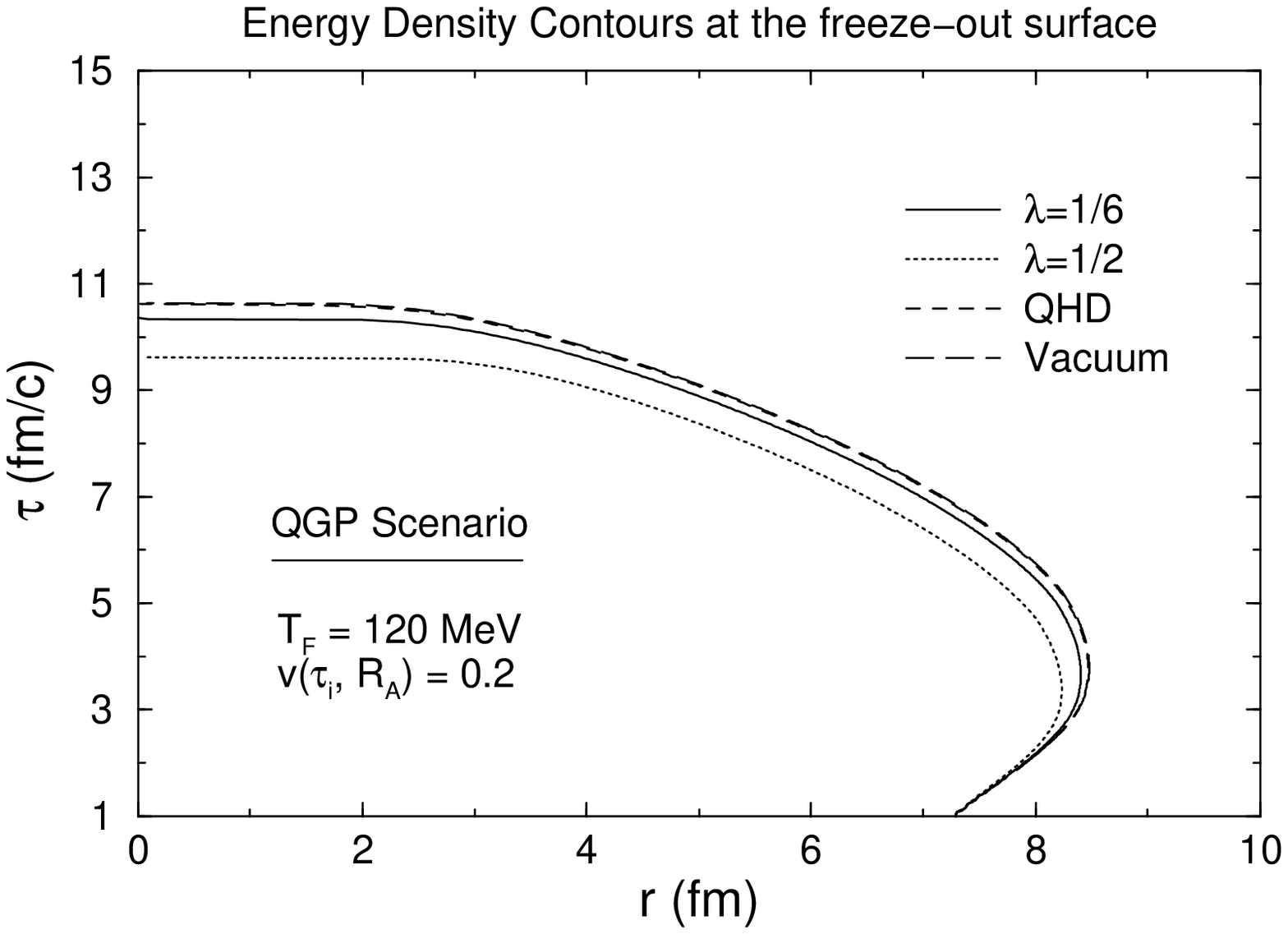,width=7.5cm}  &
   \psfig{file=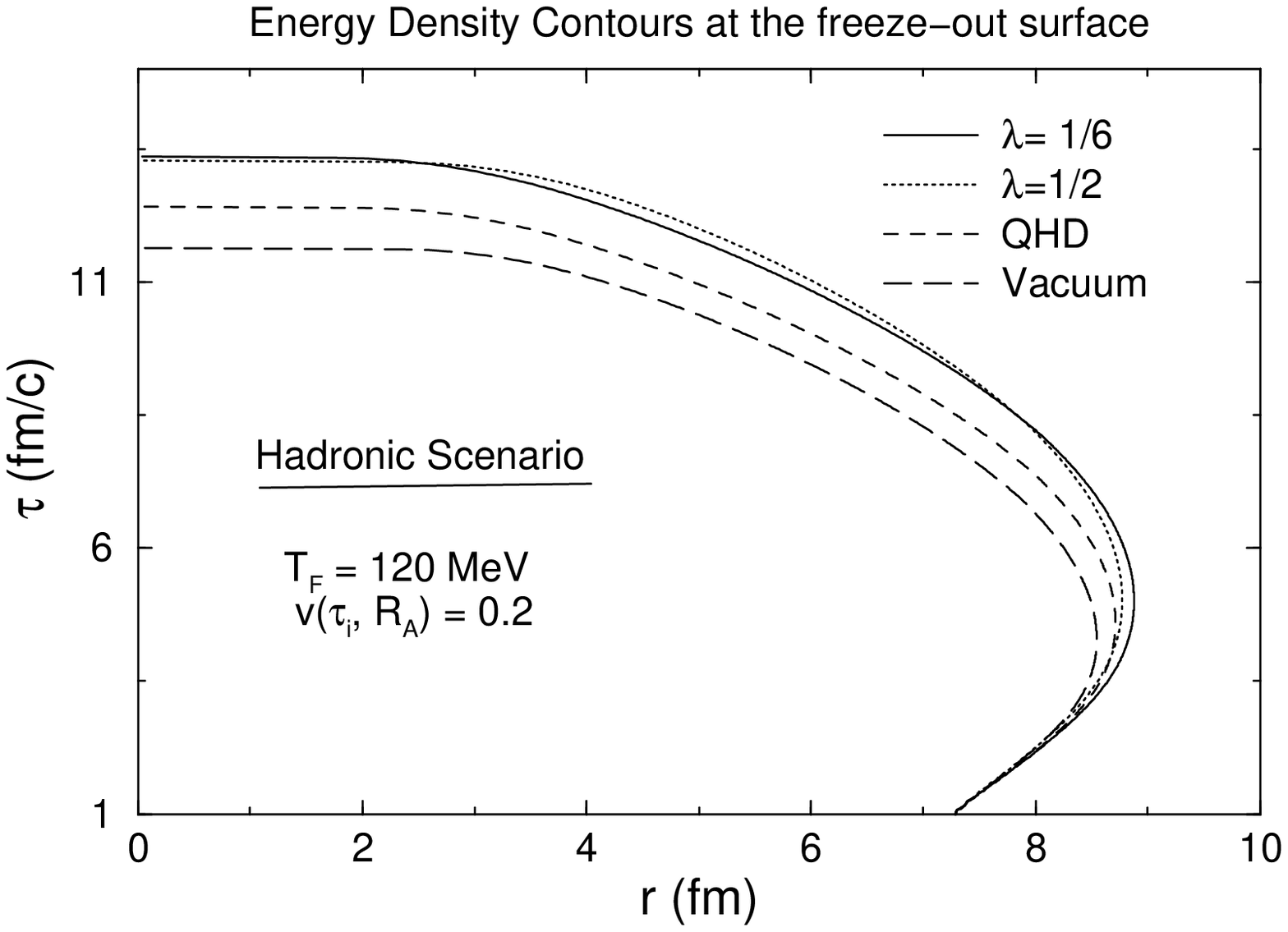,width=7.5cm}  \\
\end{tabular}
\caption{Same as fig.~\ref{fig1ab} for $v_0=0.2$.}
\label{fig2ab}
\end{figure}

In fig.~\ref{fig2ab} the results for non-zero initial radial
velocity on the surface of the cylinder is depicted.
Because of the rapid expansion of the system the freeze-out 
surface is smaller ({\it i.e.} the life time is shorter)
in this case as compared to the case 
where  the initial radial velocity is zero at the surface.
However, the qualitative shape of the curves are same as
before.

\begin{figure}
\begin{tabular}{cc}         
   \psfig{file=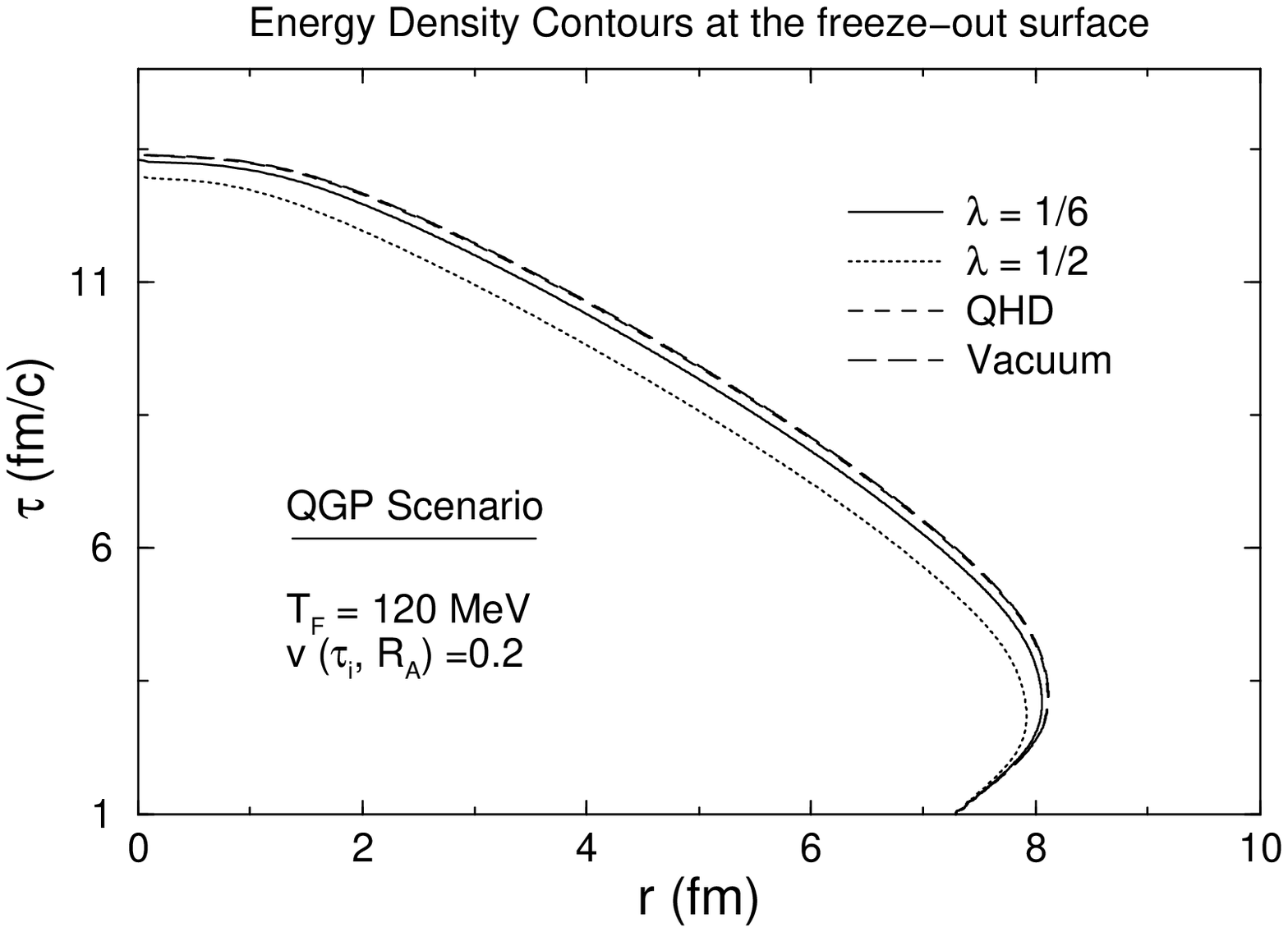,width=7.5cm}  &
   \psfig{file=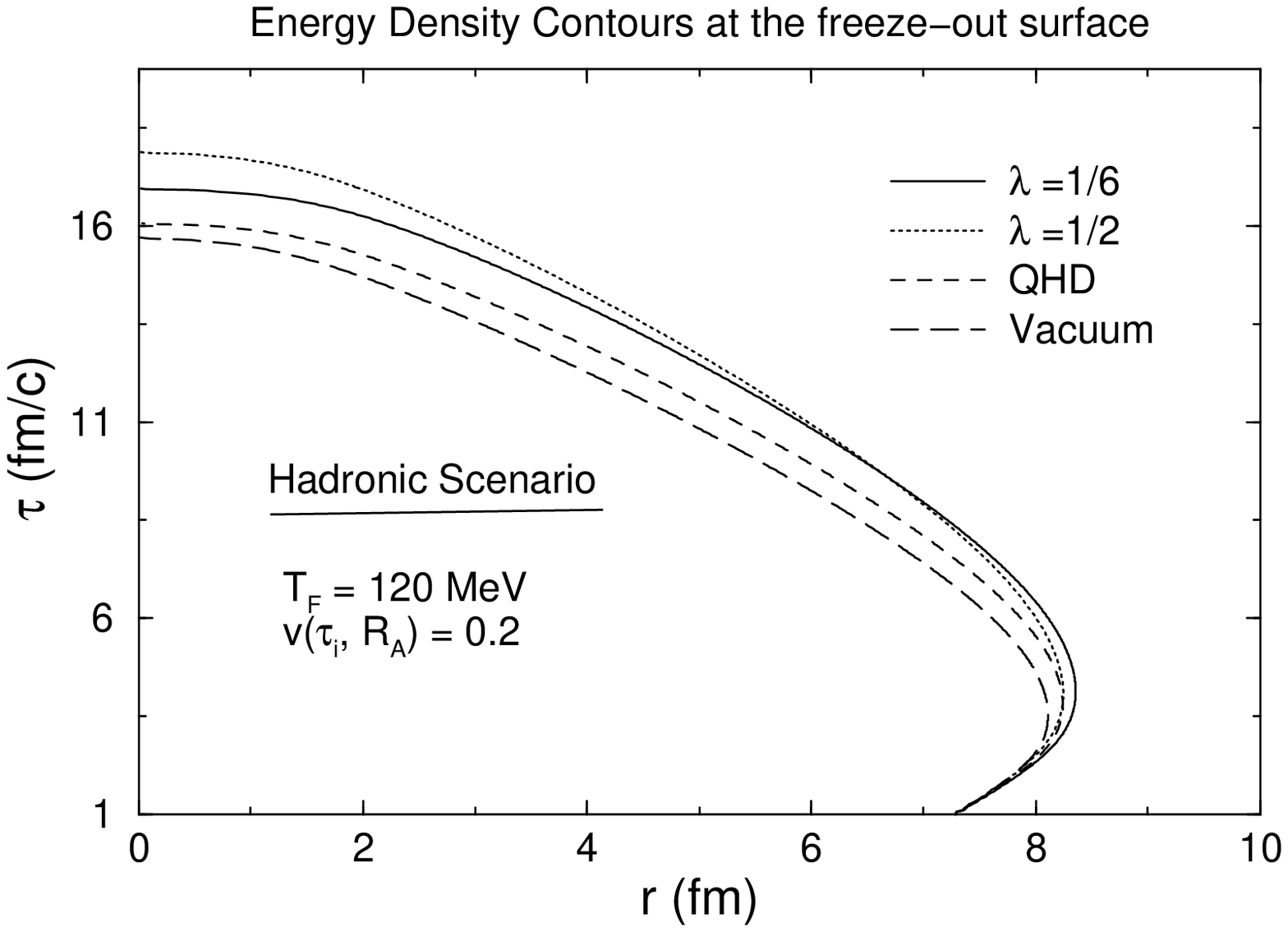,width=7.5cm}  \\
\end{tabular}
\caption{Same as fig.~\ref{fig1ab} with initial velocity
profile of Eq.~\protect\ref{v02} for $v_0=0.2$. 
}
\label{fig4ab}
\end{figure}

Fig.~\ref{fig4ab} shows the FS when a different 
initial velocity profile, 
Eq.~\ref{v02} is used with
$v_0=0.2$.  It is clear that here the expansion is 
slower than the case in fig.~\ref{fig2ab}, 
indicating that the radial velocity proportional
to radial co-ordinate gives rise to stronger flow.

\begin{figure}
\begin{tabular}{cc}         
   \psfig{file=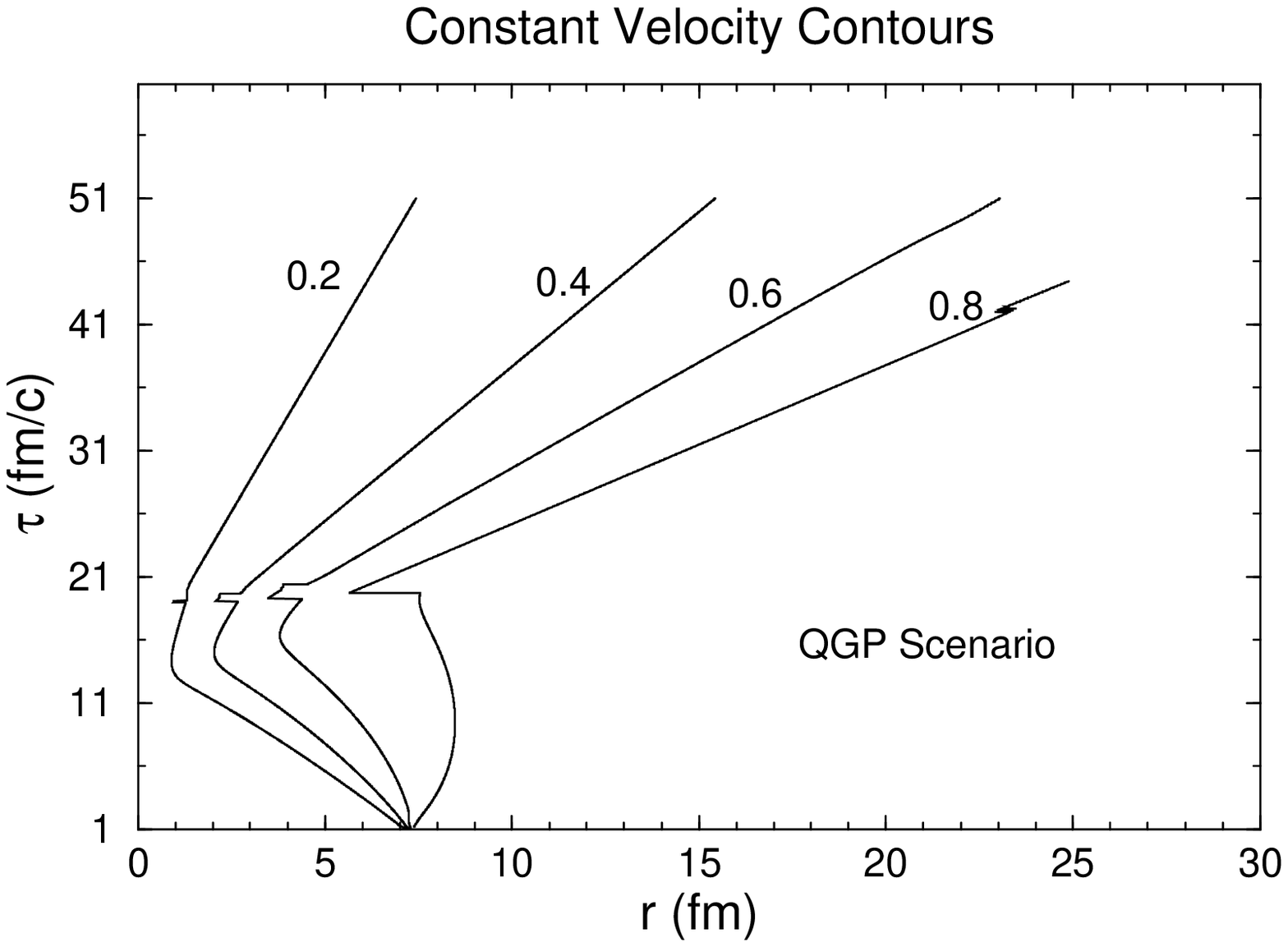,width=7.5cm}  &
   \psfig{file=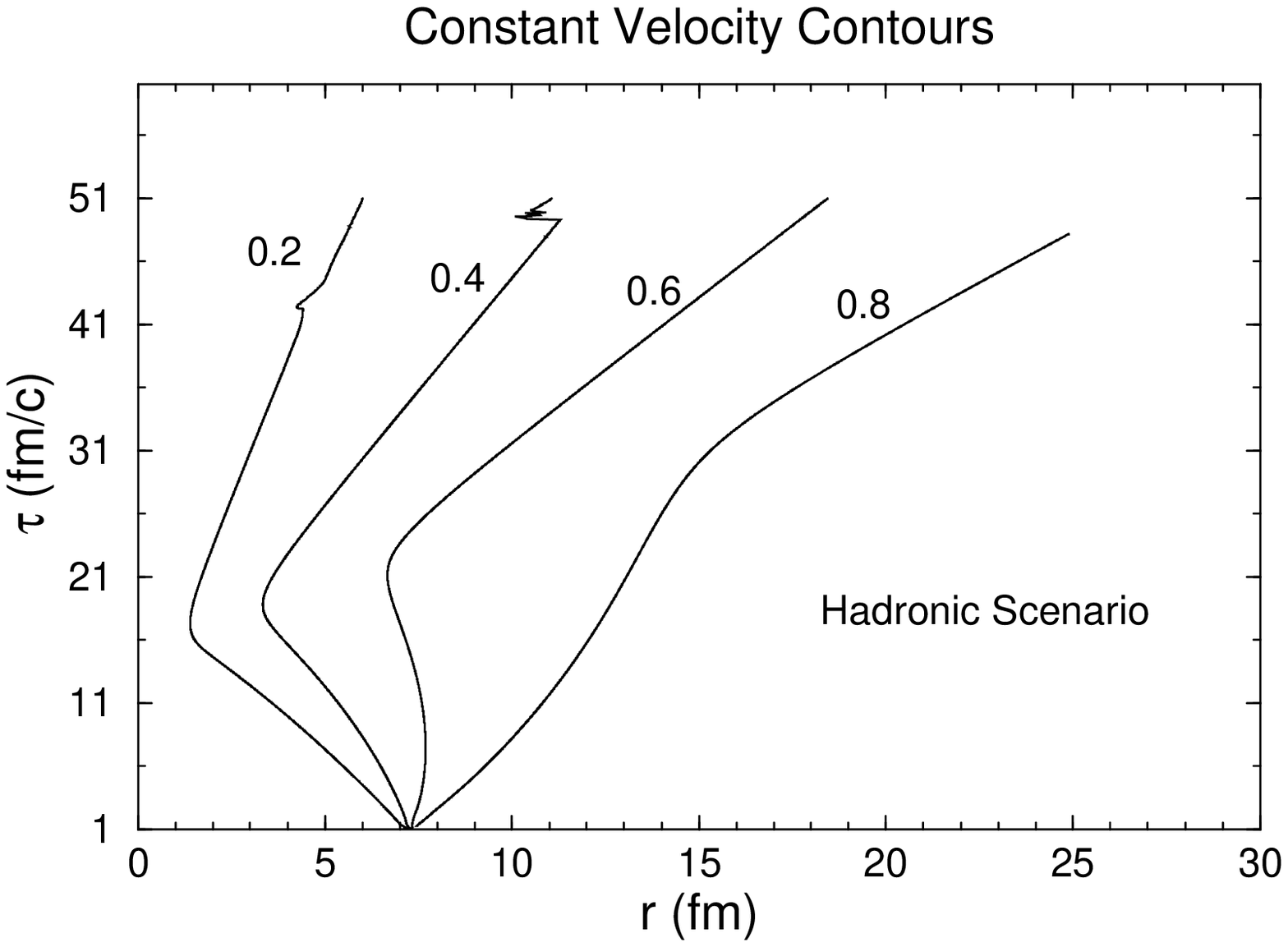,width=7.5cm}  \\
\end{tabular}
\caption{The constant velocity contours for the initial velocity
profile of Eq.~\protect\ref{v01} with $v_0=0$. Left (Right): QGP (hadronic
initial state  for vacuum masses of hadrons). 
}
\label{fig5ab}
\end{figure}

\begin{figure}
\begin{tabular}{cc}         
   \psfig{file=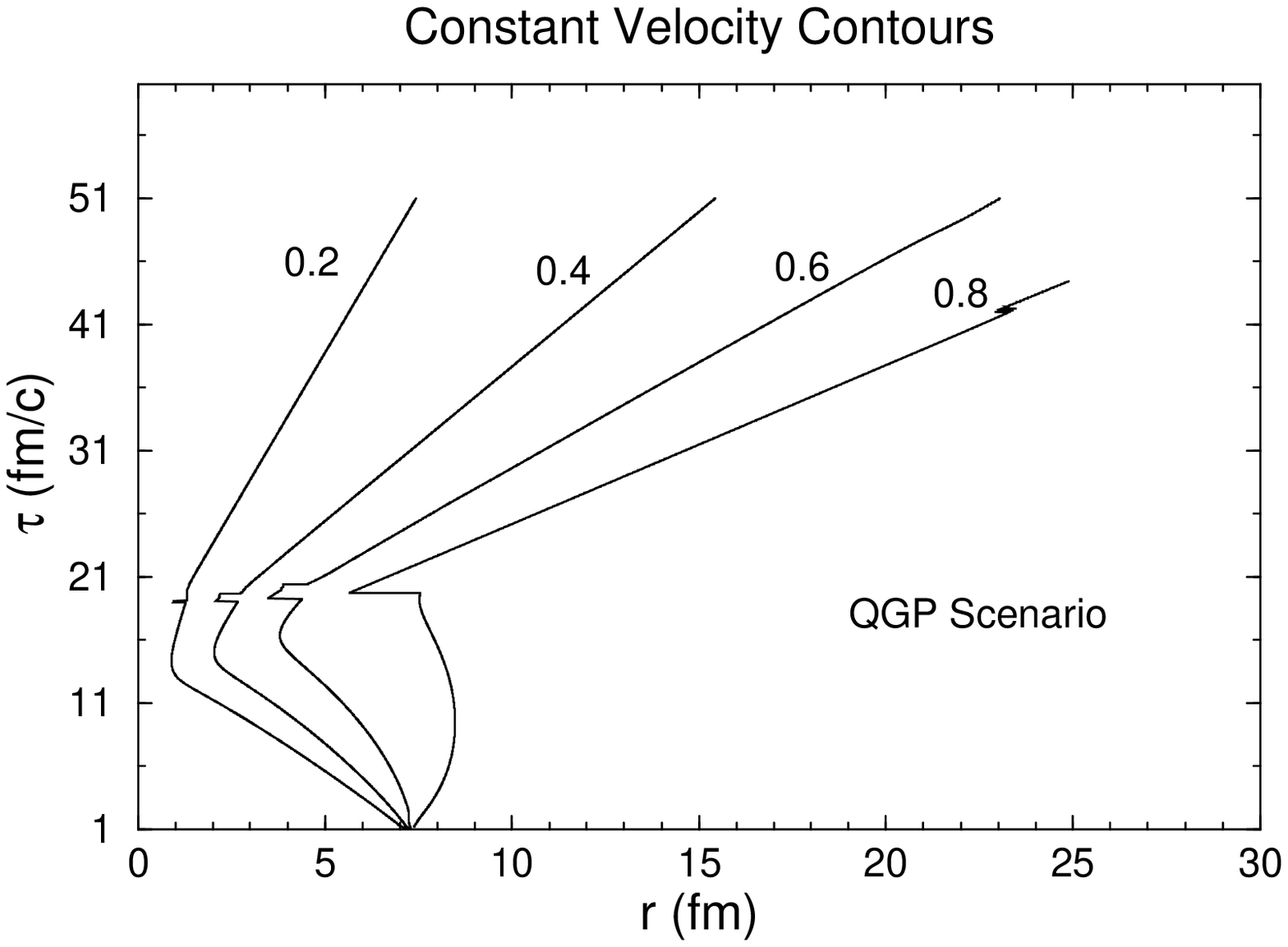,width=7.5cm}  &
   \psfig{file=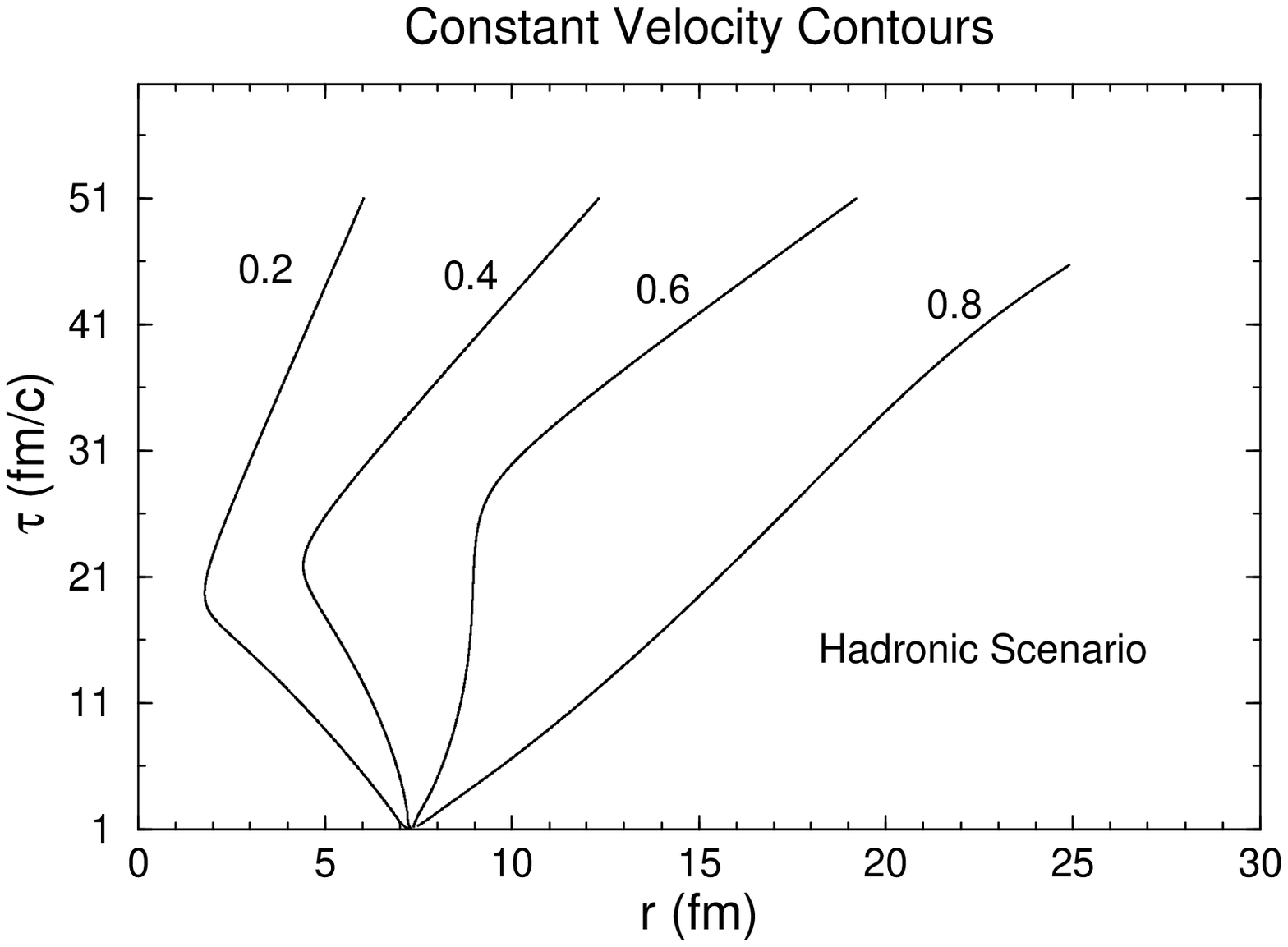,width=7.5cm}  \\
\end{tabular}
\caption{Same as fig.~\ref{fig5ab} for in-medium hadronic masses
given by Eq.~\protect\ref{emass}.
}
\label{fig6ab}
\end{figure}

\begin{figure}
\begin{tabular}{cc}         
   \psfig{file=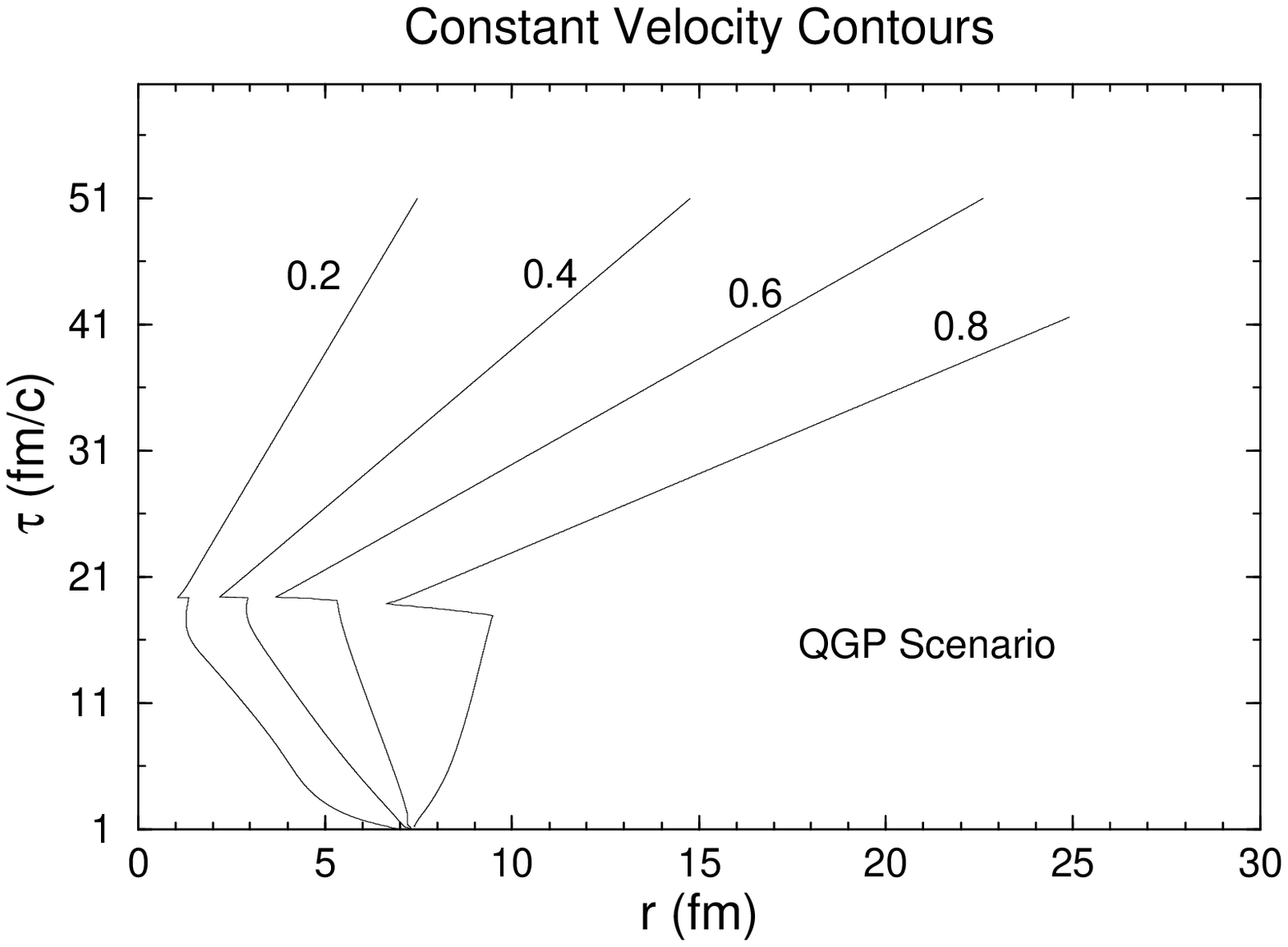,width=7.5cm}  &
   \psfig{file=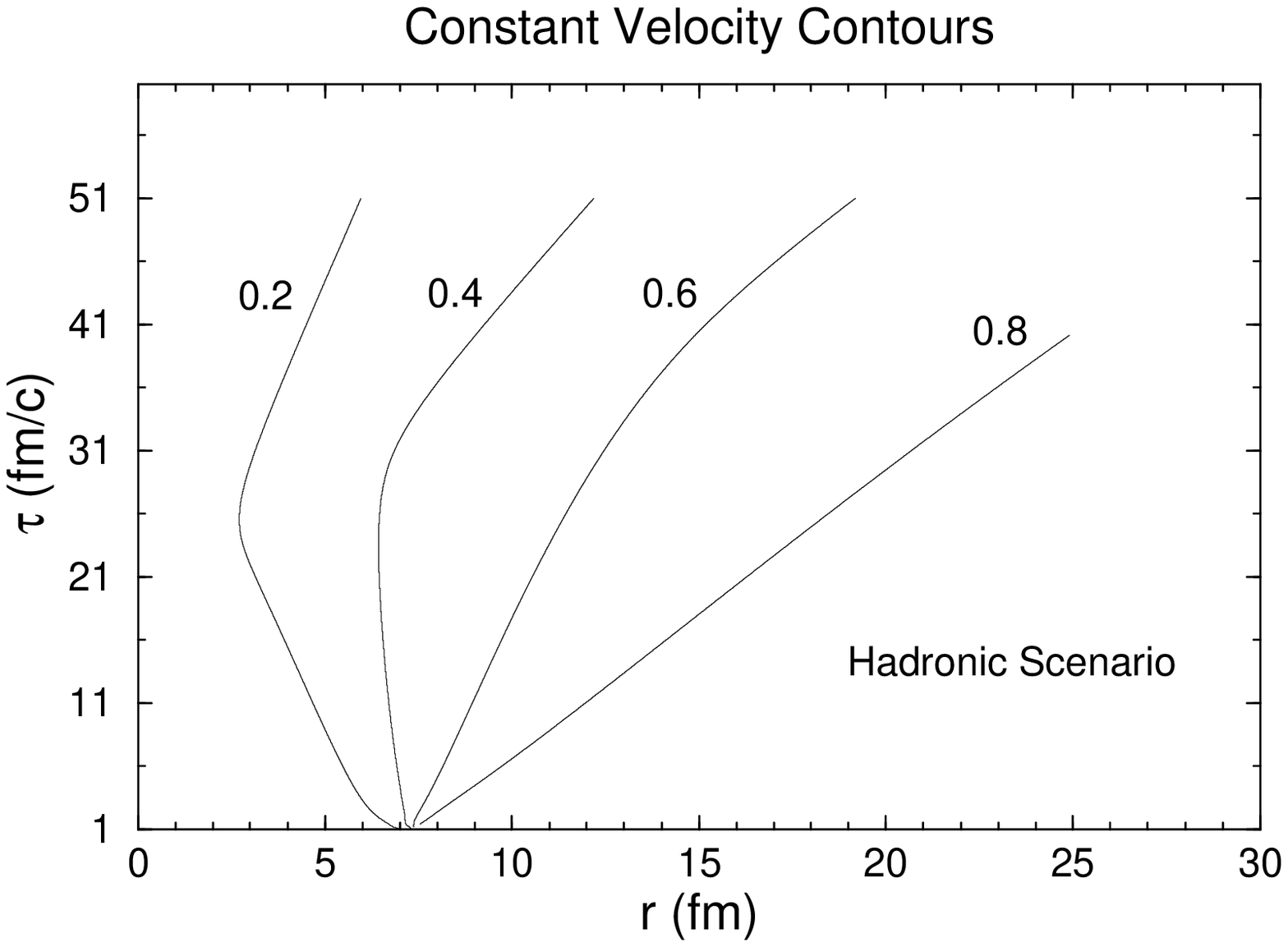,width=7.5cm}  \\
\end{tabular}
\caption{Same as fig.~\ref{fig6ab} for initial velocity
profile of Eq.~\ref{v02} with $v_0=0.2$.
}
\label{fig7ab}
\end{figure}

In fig.~\ref{fig5ab} we show the constant 
velocity contours for QGP and hadronic initial state 
when the medium effects on the hadrons are ignored. 
The initial velocity profile is given by Eq.~\ref{v01}
with $v_0=0$. Introduction of the medium effects 
does not change the qualitative behaviour of the
contours as shown in fig.~\ref{fig6ab}. In case of the QGP formation
scenario the introduction of the medium effects has
negligible effects on the space time evolution of the
system (left panel of figs.~\ref{fig5ab} and ~\ref{fig6ab}).
However, in case of the hadronic scenario 
(right panel of figs.~\ref{fig5ab} and ~\ref{fig6ab})
a visible quantitative change is evident.
In fig.~\ref{fig7ab} the effects of non-zero initial
radial velocity are  shown, a quantitative difference 
with the results of the previous figure indicates 
a large transverse flow in this case. 

\begin{figure}
\begin{tabular}{cc}         
   \psfig{file=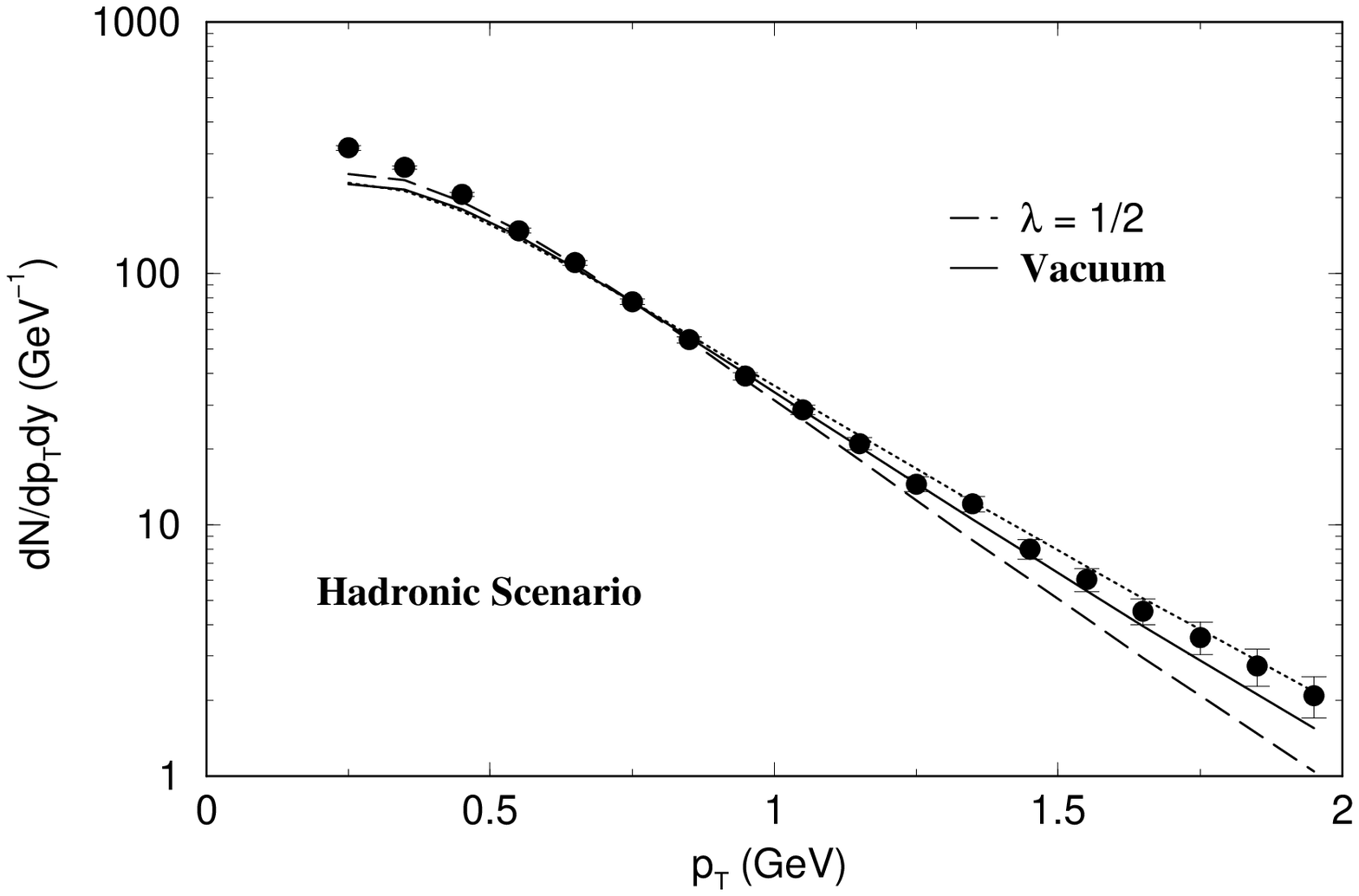,width=7.5cm}  &
   \psfig{file=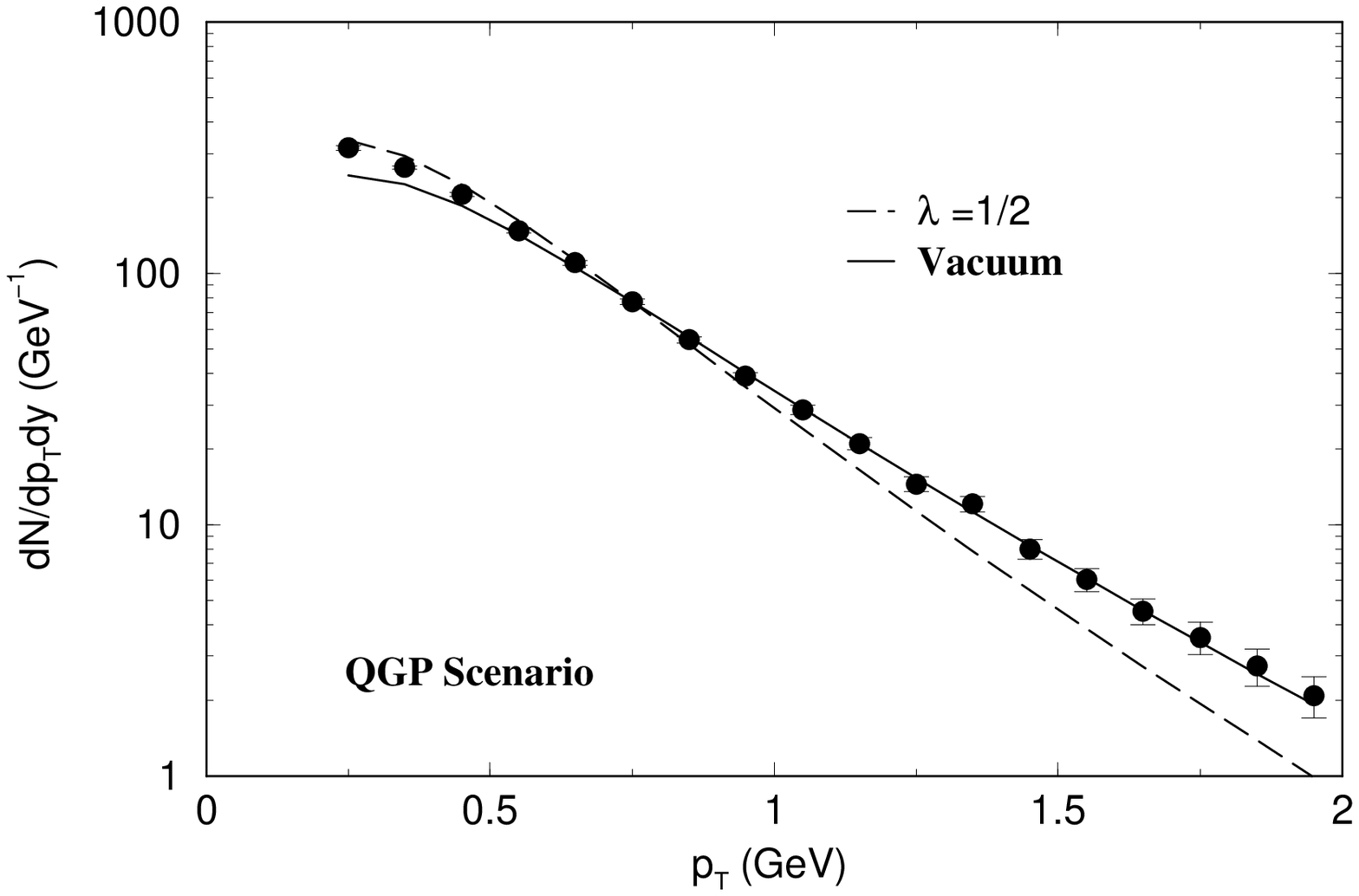,width=7.5cm}  \\
\end{tabular}
\caption{The $p_T$ distribution of pions for Pb + Pb collisions
at CERN SPS energies. Left (Right): hadronic (QGP) initial state. 
The hadronic masses vary according to 
Eq.~\protect\ref{emass} with $\lambda=1/2$. 
The initial velocity profile is taken from Eq.~\protect{\ref{v01}}
with $v_0=0$.
The dotted curve in the left panel shows the pion spectra 
evaluated for the initial radial velocity of Eq.~\ref{v01}
with $v_0=0.2$.
}
\label{fig9ab}
\end{figure}

\begin{figure}
\begin{tabular}{cc}         
   \psfig{file=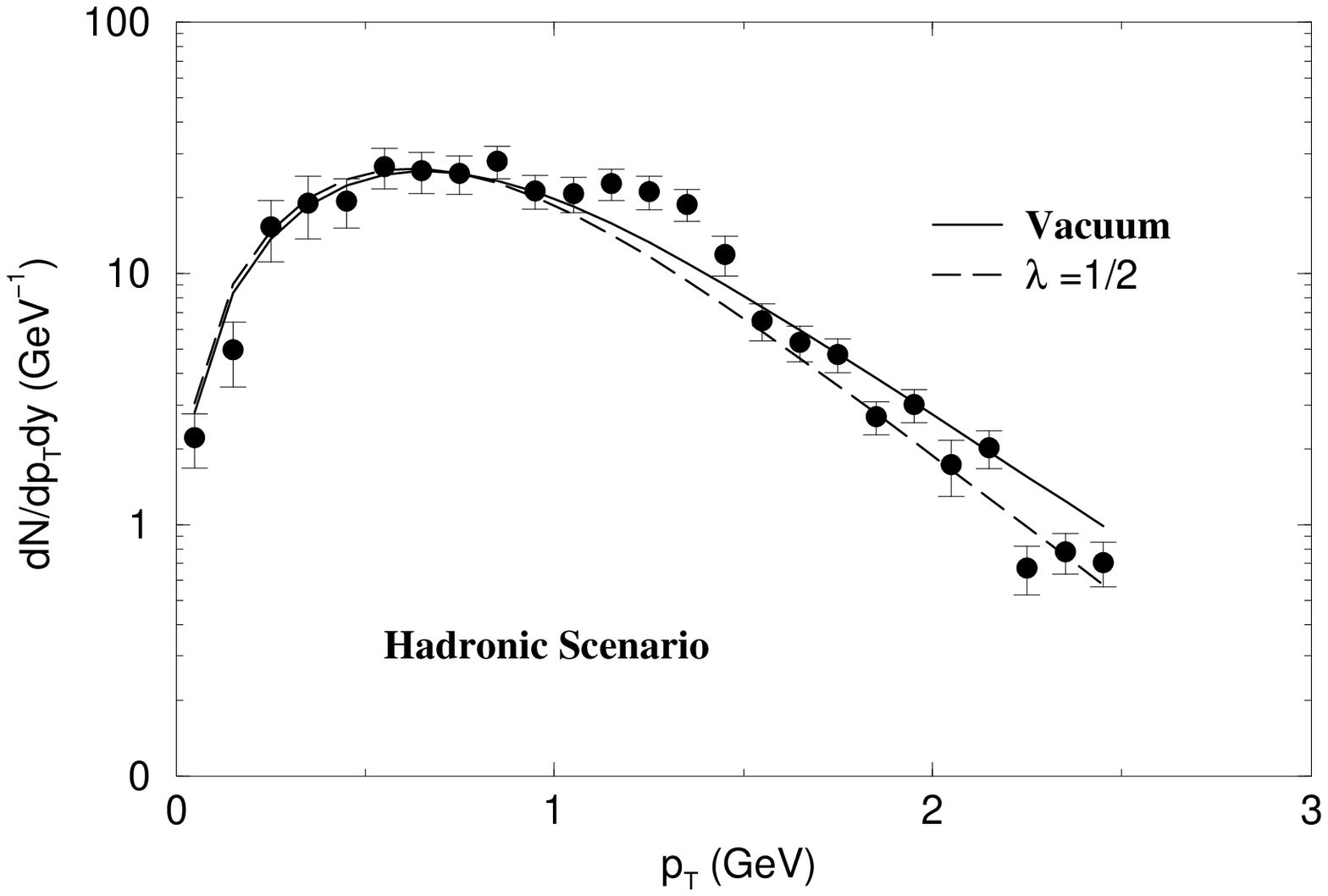,width=7.5cm}  &
   \psfig{file=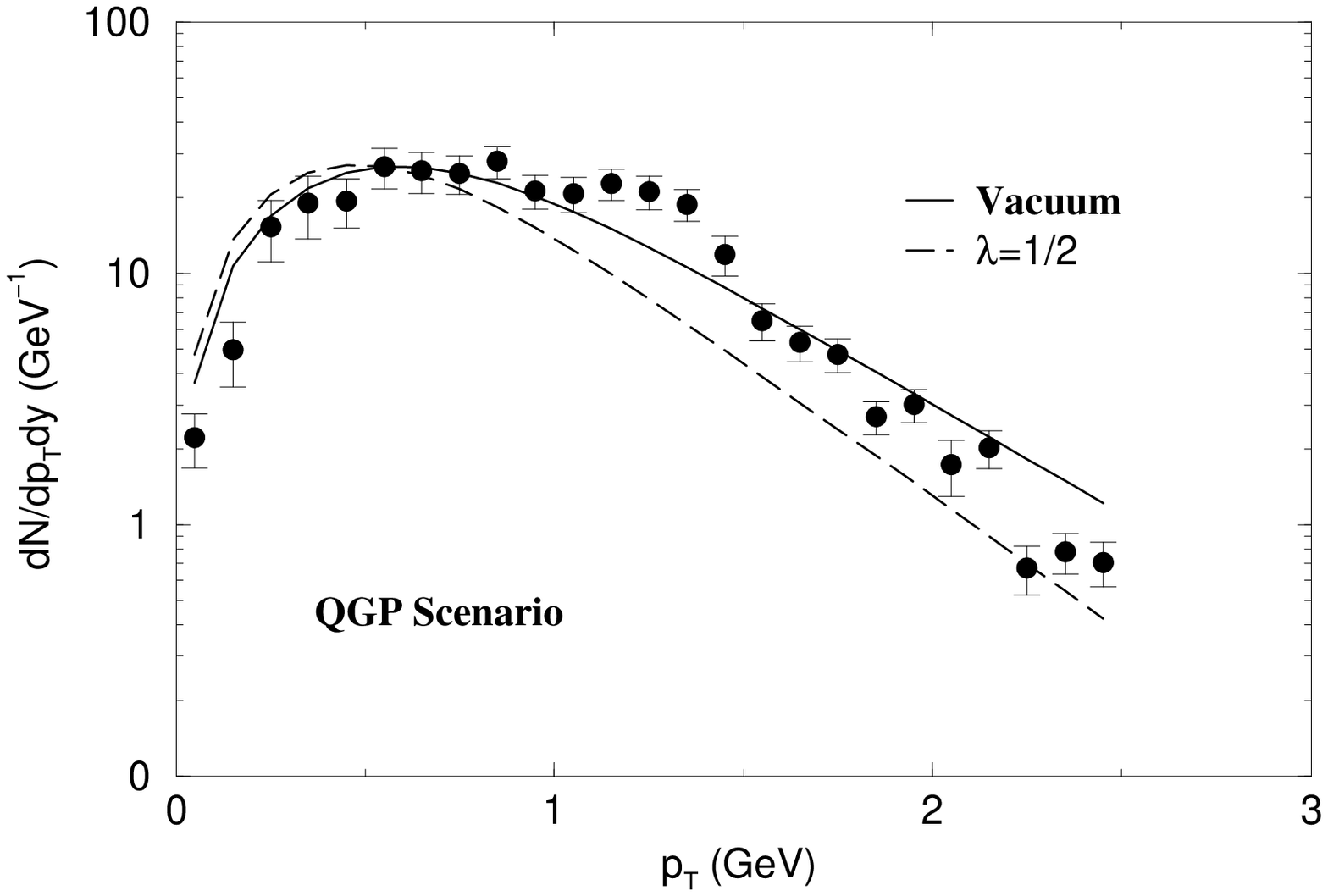,width=7.5cm}  \\
\end{tabular}
\caption{The $p_T$ distribution of protons for Pb + Pb collisions
at CERN SPS energies. Left (Right): hadronic (QGP) initial state. 
The initial velocity profile is taken from Eq.~\protect{\ref{v01}}
with $v_0=0$.
The hadronic masses vary according to 
Eq.~\protect\ref{emass} with $\lambda=1/2$.
}
\label{fig10ab}
\end{figure}

Now we discuss the $m_T$ spectra of pions and protons.
We assume that pions and nucleons are in thermal equilibrium
throughout the hydrodynamic evolution, and they freeze-out
at a common temperature $T_F=120$ MeV. 
We recall that there is no unique description so far for the
$p_T$ spectra  of hadrons.  In~\cite{wang} it has been shown   
that the $\pi^0$ spectra of WA80 and WA98 can be reproduced
by perturbative QCD calculation if the initial $p_T$ broadening
is taken into account. Wang~\cite{wang} has argued that the high $p_T$ 
$\pi^0$ spectra in central Pb + Pb collisions can not be
due to collective flow. The $p_T$ broadening in heavy ion
collisions can also be described reasonably well by a random
walk model~\cite{leonidov} where transverse flow effects are
not required to reproduce the data. Recently it has been
proposed that the $p_T$-broadening in high energy nuclear
collisions can be generated by the Color Glass Condensate
i.e. by the initial partonic phase formed after the
collisions~\cite{bielich}. Keeping these possible 
scenarios of $p_T$ broadening in mind we do not
attempt to reproduce the absolute normalization of the
high $p_T$ part of the
pion spectra by hydrodynamic flow. We rather  
concentrate on the effects of the EOS containing
the in-medium mass modification of hadrons on the 
$p_T$ spectra of pions and protons. 
We treat the normalization
of pions and nucleons as parameters to be determined by the
experimentally measured spectra. Negative hadrons and positive
minus the negative are treated as pions and protons respectively. 
In fig.~\ref{fig9ab} the NA49 pion spectra is compared  with
both QGP and hadronic initial states. 
The hadrons ($\pi^-$ and $K^-$) from the resonance decays
($\rho$ and $\phi$) are about $5\%$ of the direct pions
at $p_T\sim 500$ MeV and are therefore neglected here.
Results for QGP (hadronic)
initial state is displayed in the left (right) panel in fig.~\ref{fig9ab}.
Both the initial states describe the data reasonably well for low
transverse momentum. 
In fig.~\ref{fig10ab} the similar results are shown for protons.
The experimental spectra is well reproduced by 
the hadronic as well as QGP initial state. 
However, due to various uncertainties which are discussed
above it is not possible to state which of the EOS
is realized in such collisions.

In fig.~\ref{fig11} the $p_T$ spectra for negative hadrons at RHIC
energies are shown 
for the same value of the freeze-out temperature ( 120 MeV) and the
EOS as shown in fig.~\ref{fig1}. The agreement of our calculation
with the experimental data ~\cite{STAR} is reasonably well.
In this case a thermalized QGP with initial temperature $\sim$ 300 MeV
and thermalization time 0.5 fm/c is assumed. 

%%%%%%%%%%%%%% Fig.12 %%%%%%%%%%%%%%%%%%%%%%%%%%%%%
\bef
\centerline{\psfig{figure=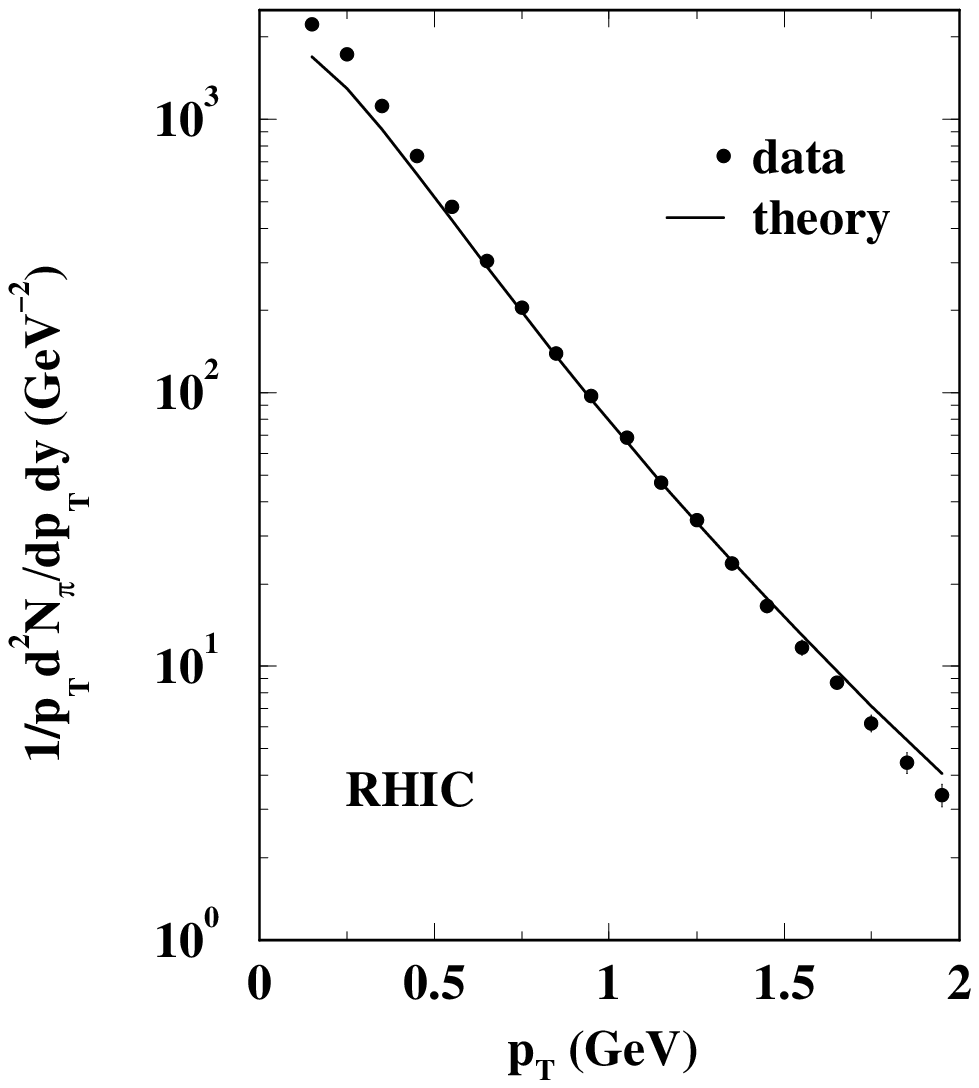,height=5.5cm,width=8cm}}
\caption{The $p_T$ distribution of the pions  at RHIC energies.
The initial velocity profile is taken from Eq.~\protect{\ref{v01}}
with $v_0=0$.
The theoretical results are obtained for  QGP
initial state at a temperature 300 MeV. 
}
\label{fig11}
\eef
%%%%%%%%%% End of Fig. 12 %%%%%%%%%%%%%%%%%%%%%%%%%%%%%%%%

\section{Summary and discussions}
We have solved the (3+1) dimensional hydrodynamical
equations for QGP as well as hadronic initial states.
The shift of hadronic masses at non-zero temperature
are incorporated through the EOS. The FS is seen to
be modified due to medium effects depending upon the
magnitude of the shift in the hadronic masses. 
For a hadronic initial state in the universal scaling 
scenario (with the value
of the exponent, $\lambda=1/2$) the FS is larger
compared to the case when no medium effects is considered.
The difference in FS for QHD and vacuum scenario is
negligible due to small mass shift in QHD near the freeze-out
point.  The pion and proton
spectra is well reproduced by a hadronic initial state 
when the mass shift of hadrons are
taken into account via the EOS. 
It is seen that the pion and proton
spectra is well reproduced  with the same equation of state,
initial conditions and the freeze-out parameters which describe 
the WA98~\cite{jarap}
and the CERES/NA45 dilepton data~\cite{ss}, 
indicating a thermal source
of initial temperature $\sim 200$ MeV at CERN SPS energies.

\vskip 0.2in
\noindent{{\bf Acknowledgement}: 
J.A. would like to thank K. Redlich for encouragement on this work.


\begin{thebibliography}{99}

\bibitem{qm01} Proc. of the 15th Int. Conf. on 
Ultra-Relativistic Nucleus-Nucleus Collisions (Quark Matter-2001),
January 15-20, 2001, The university of New York at
Stony Brook and Brookhaven National Laboratory, USA,
to be published in  Nucl. Phys. A.  

\bibitem{hadpt1} J. Cleymans and K. Redlich, Phys. Rev. {\bf C 60}
054908 (1999). 

\bibitem{hadpt2} P. Huovinen, P. F. Kolb, U. Heinz, P. V. Ruuskanen
and S. A. Voloshin, Phys. Lett. {\bf B 503} 58 (2001). 

\bibitem{hadpt3} P. Braun-Munzinger, J. Stachel, J. P. Wessels and N. Xu, 
Phys. Lett. {\bf B 365} 1 (1996).

\bibitem{hadpt4} D. Teaney, J. Lauret and E. V. Shuryak,
nucl-th/0110037. 

\bibitem{hadpt5} J. Sollfrank, P. Huovinen, M. Kataja, P. V. Ruuskanen,
M. Prakash and R. Venugopalan, Phys. Rev. {\bf C 55} 392 (1997).

\bibitem{ceres} G. Agakichiev {\it et al}, CERES Collaboration, Phys. Lett.
{\bf B 422}, 405 (1998); B. Lenkeit, Doctoral Thesis, Universitaet
Heidelberg (1998).

\bibitem{RW} R. Rapp and J. Wambach, Adv. Nucl. Phys. {\bf 25}, 1 (2000).

\bibitem{jarap} J. Alam, S. Sarkar, T. Hatsuda, T. K. Nayak
and B. Sinha, Phys. Rev. {\ C 63} 021901R (2001).

\bibitem{wa98} M. M. Aggarwal et al., WA98 Collaboration, 
Phys. Rev. Lett. {\bf 85}, 3595 (2000). 

\bibitem{na49} H. Appelsh\"auser et al., NA49 Collaboration, 
Phys. Rev. Lett. {\bf 82}, 2471 (1999).

\bibitem{jdb} J. D. Bjorken, Phys. Rev. {\bf D 27}, 140 (1983).

\bibitem{hvg} H. von Gersdorff, M. Kataja, L. McLerran and
P. V. Ruuskanen, Phys. Rev. {\bf D 34}, 794 (1986). 

\bibitem{heinz} U. Heinz, K. S. Lee and E. Schnedermann,  in Quark
Gluon Plasma, edited by R. C. Hwa, (World Scientific, Singapore 1992).

\bibitem{pbm} P. Braun-Munzinger, J. Stachel, J. P. Wessels and
N. Xu, Phys. Lett. {\bf B 365}, 1 (1996).

\bibitem{cooper} F. Cooper and G. Frye, Phys. Rev. {\bf D 10}
186 (1974).

\bibitem{ruuskanen} P. V. Ruuskanen, Acta Phys. Pol. {\bf A 18},
551 (1986).

\bibitem{blaizot} J. P. Blaizot and J. Y. Ollitrault, in Quark
Gluon Plasma, edited by R. C. Hwa, (World Scientific, Singapore 1992).

\bibitem{brpr} G. E. Brown and M. Rho, Phys. Rep. {\bf 269}, 333 (1996);
Phys. Rev. Lett. {\bf 66} 2720 (1991).

\bibitem{hsk} T. Hatsuda, H. Shiomi and
H. Kuwabara, Prog. Th. Phys. {\bf 95} 1009 (1996).

\bibitem{npa12} S. Sarkar, J. Alam, P. Roy, A. K. Dutt-Mazumder,
B. Dutta-Roy and B. Sinha, Nucl. Phys. {\bf A634}, 206 (1998);
P. Roy, S. Sarkar, J. Alam and B. Sinha,
Nucl. Phys. {\bf A 653}, 277 (1999). 

\bibitem{sxk} C. Song, P. W. Xia and
C. M. Ko, Phys. Rev. {\ bf C52} 408 (1995).

\bibitem{weinberg} S. Weinberg, Phys. Rev. Lett. {\bf 18} 507 (1967).

\bibitem{lattice} F. Karsch, hep-lat/0106019 (2000).

\bibitem{hb} G. E. Brown, H. A. Bethe, A. D. Jackson and P. M. Pizzochero,
Nucl. Phys. {\bf A 560}, 1035 (1993).

\bibitem{wang} X. N. Wang, Phys. Rev. Lett. {\bf 81}, 2655 (1998).

\bibitem{leonidov} A. Leonidov, M. Nardi and H. Satz, Z. Phys. {\bf C 74} 
535 (1997); J. Alam, J. Cleymans, K. Redlich and H. Satz, nucl-th/9707042. 

\bibitem{bielich}  L. McLerran and J. Schaffner-Bielich, Phys. Lett.
{\bf B514} 29 (2001); J. Schaffner-Bielich, D. Kharzeev, 
L. McLerran and R. Venugopalan, nucl-th/0108048.

\bibitem{STAR} C. Adler et al. (STAR Collaboration), Phys.
Rev. Lett. {\bf 87} 112303 (2001). 

\bibitem{ss} S. Sarkar, J. Alam and T. Hatsuda, nucl-th/0111032.

\end{thebibliography}
\end{document}